# Challenges for automated spike sorting:

# beware of pharmacological manipulations

Gerrit Hilgen

Biosciences Institute, Faculty of Medical Sciences, Newcastle University,

Newcastle, NE2 4HH, United Kingdom

The advent of large-scale and high-density extracellular recording devices allows simultaneous recording from thousands of neurons. However, the complexity and size of the data makes it mandatory to develop robust algorithms for fully automated spike sorting. Here it is shown that limitations imposed by biological constraints such as changes in spike waveforms induced under different drug regimes should be carefully taken into consideration in future developments.

Extracellular probes provide an excellent tool for long-term *in vitro* and *in vivo* recordings of neurons. The variety of these devices spans from single electrodes (1 channel) over tetrodes (4 channels) to multielectrode arrays (MEAs) and probes (up to thousands of channels). These devices usually pick up the activity of multiple neurons on each channel, which necessitates spike sorting to isolate the responses of single-units. For devices with few and well-separated channels, it is standard to detect and extract the spike waveforms and combine dimensionality reduction of the waveforms with posthoc clustering (for review, Rey, Pedreira, and Quian Quiroga 2015), either in a supervised or semisupervised manner. For the newest generation of CMOS (complementary metal-oxide-semiconductor-based) MEA devices with thousands of closely spaced channels, however, the sheer volume combined with the intricacy of the recordings makes it mandatory to minimize manual intervention and to develop new fully-unsupervised spike sorting workflows (for review, Hennig, Hurwitz, and Sorbaro 2018). These new algorithms not only need to tackle the computational challenges but they also need to consider acute changes in the spike waveform caused by overlapping spikes, drifting electrodes or bursting activity (for review, see Rey, Pedreira, and Quian Quiroga 2015). In addition, acute changes can occur in spike waveforms in the presence of pharmacological agents that affect ionic conductances. Such variability, if not taken into consideration, can impair the quality of automated spike sorting, and consequently, bias the biological interpretation of the results. Recently developed unsupervised spike sorters (see Hennig, Hurwitz, and Sorbaro 2018 for a complete list) use either template-matching or density-based approaches to overcome these challenges. For template-matching the cluster size is set lower than estimated, ending up with multiple clusters from the same unit. These clusters are then iteratively compared to each other and merged if they represent the same neuron (for review, see Lefebvre, Yger, and Marre 2016). This approach works very well for very short-term spike waveform changes caused by changes in activity levels. However, for longer-term changes (over the course of minutes to hours) induced by direct or indirect modulation of ion channel properties under different pharmacological and chemogenetic regimes, the template information provided by these affected waveforms would be falsely interpreted as spikes originating from different neurons, requiring manual intervention to reassign the waveforms to the correct neurons. On the other hand, density-based algorithms (Chung et al. 2017; Hilgen et al. 2017; Jun et al. 2017) are able to produce reliable results in such conditions, without necessitating manual intervention.

This review summarizes the different ionic components of the spike waveform, lists conditions which can affect these components either acutely or chronically, and gives a brief overview of the existing unsupervised density-based spike sorters.

*The spike waveform components:* Considering the classical Hodgkin Huxley model of axonal spikes, a spike waveform is composed of the initial resting membrane potential phase (Fig. 1A, 1) followed by a small change of the membrane potential to the point where it crosses the threshold potential. At that point, voltage-gated $Na^+$ channels open, resulting in a massive $Na^+$ influx into the neuron (*depolarization*, Fig. 1A, 2). After about a very short period of threshold depolarization, these same $^+$ channels inactivate, resulting in the repolarization phase of the action potential towards resting levels (determined by the high $K^+$ leakage permeability). At that point, the delayed-rectifier voltage-gated $K^+$ channels open, increasing $K^+$ efflux and rapid *repolarization* (Fig. 1A, 3). Until the delayed rectifier channels close again, the permeability to $K^+$ is higher than in resting conditions, causing the membrane potential to move closer to the equilibrium potential for $K^+$ (undershoot or *afterhyperpolarization*, Fig. 1A, 4). Somatic spikes, however, have more than a dozen distinct voltage-dependent conductances, and ionic channel contributions to the spike waveform are much more complex than in the axon (for review, Bean 2007).

*Pharmacology-induced changes in spike waveforms:* Obviously, applying various voltage-gated ion channel agonists and antagonists will directly modify the spike waveform. Drugs such as tetrodotoxin (TTX, $Na^+$ channels), 4-aminopyridine (4-AP, $K^+$ channels), tetraethylammonium (TEA, $K^+$ channels) and cadmium chloride ($CdCl_2$, $Ca^{2+}$ channels) for example, have dramatic effects on various components of the spike waveform. The $Na^+$ channel opener veratridine leads to persistent activation, therefore affecting the spike waveform as well (Akanda et al. 2009). Such compounds are routinely used in intracellular, but rarely in extracellular recordings. However, many different drugs are used to probe neural function and connectivity on a larger scale in extracellular recordings. Many of these compounds can change the spike waveform. For example, strychnine, an antagonist of the inhibitory neurotransmitter glycine, is known to affect voltage-gated Na$^+$ channels as well (Reiser, Günther, and Hamprecht 1982), thereby profoundly changing the spike waveform. The antimalarial drug quinine is often used to block specific gap junctions (Cx36), but it is also a K$^+$ channel modulator, and at high concentrations (mM range) it even modulates Na$^+$ channels (Akanda et al. 2009). Indeed, we recorded spikes with an HD MEA from the mouse ganglion cell layer and found the effect of quinine on the average spike waveform to be severe, even at 100 µM concentration (Fig. 1A, blue vs beige). The GABA$_A$ receptor antagonist bicuculline also blocks Ca$^+$-activated K$^+$ channels (Bruening-wright, Adelman, and Maylie 1999) which are responsible for the slow afterhyperpolarization phase in some types of neurons (Fig. 1A, 4). Mecamylamine is an antagonist of nicotinic cholinergic receptors with no known potential effects on spike waveforms. In the presence of a cocktail of 20 µM mecamylamine and 20 µM bicuculline, we observed a decrease in the amplitude of both phases of the average spike waveform amplitude (Fig. 1D, blue vs beige).

*Chemogenetic-induced changes of spike waveforms:* Neuroscientist now commonly use chemogenetic tools that alter various ionic conductances over periods of several hours, but little attention has been given so far to the underlying implications this might have for spike waveforms. For example, PSAM (pharmacologically selective actuator module, Sternson and Roth 2014) is a chimeric chloride-permeable ligand-gated channel for neuron activation and silencing that causes changes in membrane potential, thereby affecting spike waveforms. Further, inhibitory hM4Di (human M4 muscarinic receptor) DREADDs (designer receptors exclusively activated by designer drugs, Sternson and Roth 2014) activated by clozapine N-oxide induce hyperpolarization by activation of GIRK (G-protein inwardly rectifying K$^+$ ) channels and excitatory hM3Dq (human M4 muscarinic receptor) DREADDs induce tonic depolarization of the cell membrane by intracellular calcium release. The extent of membrane potential changes has a direct effect on the amplitude of the spike waveform, with smaller spikes when the membrane is depolarized, and larger spikes when the membrane is hyperpolarized (Nádasdy Z et al. 1998).

*Density-based algorithms:* Nádasdy et al. (1998) described a silicon probe with a densely-spaced and diamond-shaped recording site to map the isopotential contours of a spontaneously spiking rat pyramidal cell. By advancing the probe in small steps in the vicinity of the cell while measuring the amplitudes of the successively recorded spikes, it was possible to interpolate the barycenter (center of mass) of the spike amplitude. They stated that it is feasible to determine the location of the action potential generation site by using three or more recording sites with less than 50 µm spacing. Further, they also estimated the spike origins of rat cortical recordings made with a high-density 8 x 16 shank probe array on a larger scale. Briefly, spikes were detected for each channel and a weighting factor, determined by the interelectrode distance, was assigned to each measured spike amplitude. The mean of the weighted amplitudes is then plotted as a function of the sum of the amplitudes as a two-dimensional density map which can be very efficiently clustered. Indeed, Nádasdy et al. used K-Means clustering to isolate potential single units. This study laid the foundation of how to use spatial spike information for spike sorting, an approach that has recently been revived and massively improved by density-based unsupervised spike sorters (MountainSort: Chung et al. 2017; Herding Spikes: Hilgen et al. 2017; JRCLUST/IRONCLUST: Jun et al. 2017). In high-density arrays and probes with less than 60 μm spacing, spike events from the same neuron are recorded on multiple channels, a fact that is used to estimate the barycentre from amplitudes in adjacent channels. As mentioned above, the two-dimensional spacing of spike localisations can be very efficiently clustered but it is mandatory to combine that information with the information from extracted spike waveform features obtained through dimensionality reduction for successful isolation of single-units. The current density-based sorters use different clustering strategies to solve this task: JRCLUST/IRONCLUST uses a newly developed density-based clustering algorithm named DPCLUS and identifies local density peaks by calculating pairwise distances between neighboring points. MountainSort is also using a newly developed non-parametric density-based algorithm named ISO-SPLIT, which generates unimodal clusters by using isotonic regression. Herding Spikes is using the mean-shift algorithm to "herd" data points towards high-density areas.

What these density-based sorters have in common is the fact that they use the spatial origin of the spike waveform as a foundation for their sorting. This is a very important point because the spike localization remains unchanged in control and drug conditions, even when the extracted spike waveform principal components change under different drug regimes. To further illustrate this crucial issue, we plotted the first two whitened principal components of control (beige) and drug condition (blue) from the experiments mentioned above in their feature space (Fig. 1, C & F). We did not cluster these points but instead colored them according to their condition origin. It is already visible by eye that the two conditions form separate clusters, even though originating from the same neuron, and conventional clustering based on these principal components alone would likely fail. In contrast, we plotted the interpolated spike localization for all these waveforms (Fig. 1 B & E) and, again, colored them according to their condition origin. The spatial origin of the spike waveform remains unchanged in control and drug conditions and these example units were sorted correctly, unsupervised and in near real-time, with Herding Spikes.

The examples in Figure 1 are rather extreme and for most conditions, template-matching and density-based sorters perform equally well (https://spikeforest.flatironinstitute.org/). However, both sorting strategies have their limitations. Template-matching sorters often require manual curation and do not perform well on recordings with low firing rates, whereas density-based sorters require relatively uniform and dense spacing of electrodes. For data that is suitable for both types of sorters, it is advisable to use more than one sorter algorithm. A very convenient approach is to use SpikeInterface (http://homepages.inf.ed.ac.uk/mhennig/news/spikeinterface/), a newly developed spike sorting workflow to sort the same data with different sorters and compare the results in a comprehensive and standardized way in a GUI.

Acknowledgments: I thank Evelyne Sernagor (Newcastle University, UK) and Matthias Hennig (University of Edinburgh, UK) for helpful discussions and critical readings of this manuscript. Experiments were performed in Evelyne Sernagor's lab and funding was provided by the Leverhulme Trust grant RPG-2016-315 (ES).

Figures

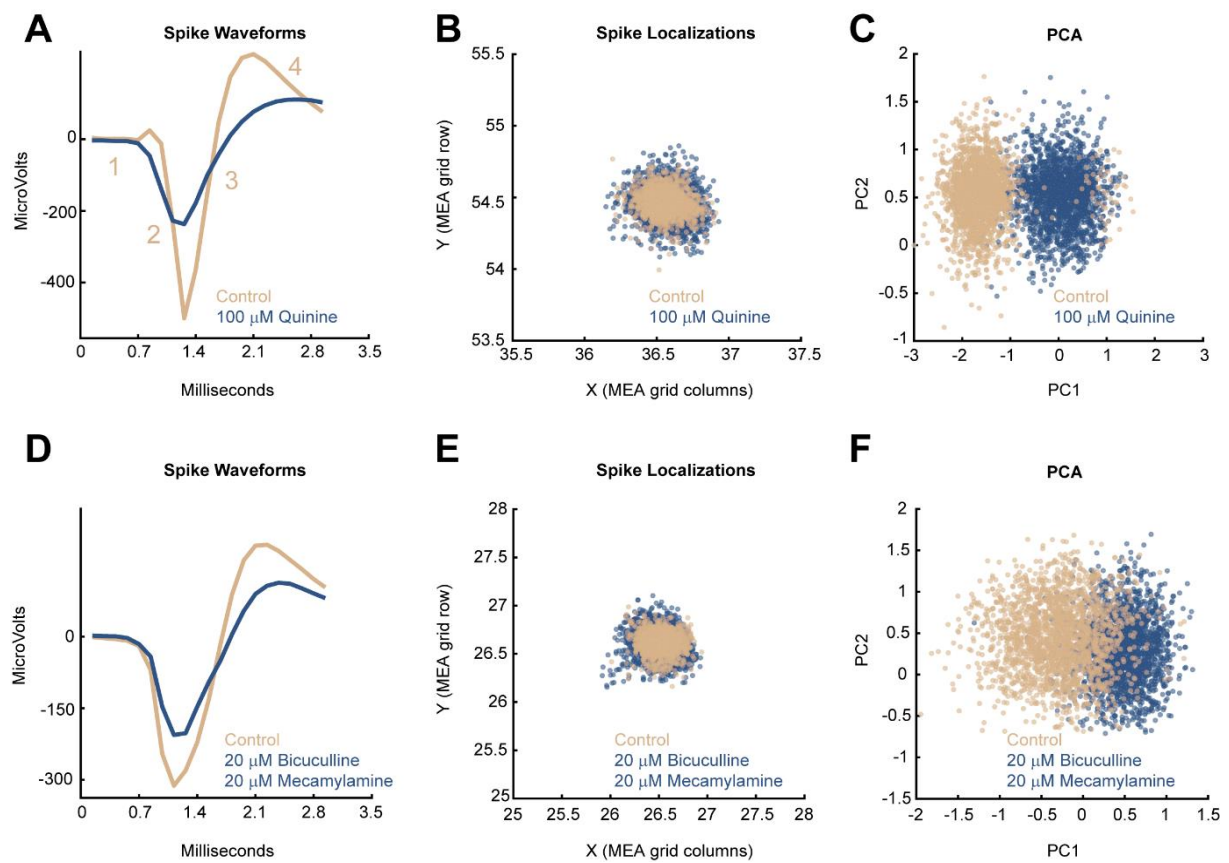

Figure 1: Drug-induced changes of the spike waveforms from mouse retinal ganglion cells recorded with an HD MEA (3Brain, Switzerland). A) The average spike waveform of a representative selected RGC in control (beige) and after application of 100μM quinine (blue). The average spike waveform is profoundly changed in the presence of quinine. Numbers indicate the different stages of a spike waveform as explained in the main text. As a convention, extracellular spikes such as those illustrated here have opposite polarity to intracellular spikes. B) The spatial origin of the spike source, the spike localization remains similar in both conditions. C) Plotting the first two principal components (Python scikit-learn module, sklearn.decomposition.PCA, whiten=true) from control and 100μM quinine spike waveforms shows two very well-separated clusters. It is difficult to assign one cluster in principal component space for this single RGC unit. D-F) Another example of drug-induced changes of spike waveforms is shown here by using 20μM bicuculline in combination with 20μM mecamylamine. D) The average waveform changes after application of this drug cocktail but the spike localization remains the same (E) while the feature space shows two clusters.